\documentclass[12pt]{iopart}

\usepackage{graphicx}
\usepackage{epstopdf}
\begin{document}

\title[Trapped BEC phase diagram]{BEC phase diagram of a $^{87}$Rb trapped gas in terms of macroscopic thermodynamic parameters}

\author{V Romero-Rochin$^1$\footnote{Corresponding author}, R F Shiozaki$^2$, M Caracanhas$^2$, E A L Henn$^2$, K M F Magalh\~aes$^2$, G Roati$^2$, V S Bagnato$^2$}

\address{$^1$Instituto de F\'{\i}sica, Universidad Nacional Aut\'onoma de M\'exico, Apartado Postal 20-364, 01000 M\'exico, D.F., Mexico}

\ead{romero@fisica.unam.mx}

\address{$^2$Instituto de F\'{\i}sica de S\~{a}o Carlos, Universidade de S\~{a}o Paulo, C.P. 369, 13560-970 S\~{a}o Carlos, SP, Brazil}

\begin{abstract}
We measure the phase diagram of a $^{87}$Rb Bose gas in a harmonic trap in terms of macroscopic parameters obtained from the spatial distribution of atoms. Considering the relevant variables as size of the cloud ${\cal V}$, number of atoms $N$ and temperature $T$, a novel parameter $\Pi = \Pi(N,{\cal V},T)$ is introduced to characterize the overall pressure of the system. We construct the phase diagram ($\Pi$ vs $T$) identifying new features related to Bose-Einstein condensation (BEC) transition in a trapped gas. A thermodynamic description of the phase transition based on purely macroscopic parameters, provide us with properties that do not need the local density approximation. An unexpected consequence of this analysis is the suggestion that BEC appears as a continuous third-order phase transition instead of being a second-order one.

\end{abstract}

\date{\today}

\pacs{67.85.-d, 05.70.Ce, 05.70.Fh}

\maketitle

Experiments with cold gases are normally performed in traps which result in spatially inhomogeneous samples\cite{Cornell1,Hulet1,Ketterle1,BECNaBrazil,BECRbBrazil}. In such cases,  the phase diagram for the transition from condensate to non-condensate phases becomes difficult to express using quantities that apply mostly for homogenous systems, say, diagrams in terms of pressure versus temperature. Part of the problem can be solved if one relies on a local-density approximation (LDA) where characterization of a non-uniform fluid, at a given spatial position, is obtained considering local quantities which in turn determine bulk properties. Since the phenomena observed, such as Bose-Einstein condensation (BEC), occur at a macroscopic level in a non-uniform trap, the description would be more appropriate if the overall characterization of the system could be achieved based on few thermodynamic parameters, which in turn should depend on other thermodynamic variables such as temperature, number of particles and properties of the confining potential.

Recently, several papers have explored macroscopic quantum phenomena based on measurements of thermodynamic properties. Horikoshi and collaborators\cite{Horikoshi} have used a force balance equation to create a state equation, which allow to determine local quantities of a Fermi gas at unitarity. Ho and Zhou\cite{Ho} have proposed an algorithm to obtain the phase diagram and thermodynamic quantities of the corresponding bulk system, by analyzing the density profile of trapped gases. Equivalent concepts were also used by Nascimbene et al.\cite{Nascimbene} to explore thermodynamics of a Fermi system.

In this work, we consider a different way to describe the phase diagram for the BEC transition, using a single macroscopic parameter to characterize the inhomogeneous system. This approach is based on concepts introduced in Refs.\cite{VRRPRL2005,VRRBJP,VRRPRE2008,BECNaBrazil}, and it is applied to experimentally analyze the BEC transition for a harmonic trapped Rb gas. However, the employed ideas are quite general and can be applied for any trapped system.

We start by considering the relevant macroscopic parameters to be used. The thermodynamic state of a gas trapped in a macroscopic harmonic potential is naturally characterized by the trapped number of atoms $N$, the temperature $T$ and parameters that determine the confining potential. In a uniform gas, apart from temperature and number of particles, the mechanical variable that specifies  the thermodynamic state of the system is the volume $V$ that the sample occupies. Since a gas confined in a harmonic trap does not have a well defined volume, one can use an alternative variable that yields analogous information regarding its size. Considering the harmonic potential $V_{ext}(\vec r) = \frac{1}{2} m (\vec \omega \cdot \vec r)^2$ with $\vec \omega = (\omega_x,\omega_y,\omega_z)$, it turns out that in a first approximation the size of the system is given by $r^{*3} \approx (k_BT/m\omega^2)^{3/2}$. Thus, apart from constants and other independent thermodynamic variables, this allows us to identify ${\cal V} = 1/\omega^3$ as the ``volume parameter", where $\omega^3 = \omega_x \omega_y \omega_z$. This variable plays the role of the actual volume in a harmonically trapped fluid. In fact, long ago, de Groot et al.\cite{Seldam}, and later Bagnato et al.\cite{Bagnato87} and others\cite{Dalfovo,Pethick,Yukalov}, demonstrated that ${\cal V}$ is the correct thermodynamic quantity determined by the external potential, in the sense that the thermodynamic limit is obtained when $N \to \infty$, ${\cal V} \to \infty$ but $N/{\cal V} \to$ constant. The volume parameter, which should not be identified with the real volume of the system, does have equivalent properties to the spatial volume of a uniform system. Among the most relevant ones is its extensive thermodynamic quality\cite{VRRPRL2005,VRRBJP,VRRPRE2008}.

Therefore, we take the collection $(N,T,{\cal V})$ as the independent thermodynamic variables that specify the state of the system. Any other thermodynamic property of the gas can be expressed as a function of those variables. That is, just as in a homogeneous fluid the phase diagram may be expressed in terms of the equation of state $p = p(N,T,V)$, with $p$ the hydrostatic pressure, here, because the state is given in terms of the ``volume parameter", we also define a ``pressure parameter" $\Pi$, such that the equation of state becomes now $\Pi = \Pi(N,T,{\cal V})$. Analogously to the volume parameter, $\Pi$ is not an actual hydrostatic pressure, although it does retain characteristics of a pressure, as we argue now. By considering Helmholtz free energy $F$ as $F = F(N,T,{\cal V})$, one can introduce $\Pi$ as the variable conjugate to ${\cal V}$, namely,
\begin{equation}
\Pi = - \left( \frac{\partial F}{\partial {\cal V}} \right)_{N,T} .\label{pi}
\end{equation}
We note that while $\Pi$ neither has units of pressure nor ${\cal V}$ units of volume, the product $\Pi {\cal V}$ still has units of energy. Clearly, $\Pi$ is an intensive thermodynamic variable and, therefore, its dependence on $(N,T,{\cal V})$ must be of the form, $\Pi = \Pi(N/{\cal V},T)$. Thus, in keeping with the analogy, we shall call $N/{\cal V}$ the density parameter. From the above definition of $\Pi$, without resorting to LDA, a very useful and general expression can be derived, this is\cite{VRRPRL2005,VRRBJP,VRRPRE2008},
\begin{equation}
\Pi = \frac{2}{3{\cal V}} \int n(\vec r) \> \frac{1}{2} m (\vec \omega \cdot \vec r)^2 \> d^3 r ,\label{piform}
\end{equation}
where $n(\vec r)$ is the inhomogeneous density profile of the gas. Since both $n(\vec r)$ and the confining potential are measurable quantities, $\Pi$ can be readily determined. With the equation of state $\Pi = \Pi(N/{\cal V},T)$,  the phase diagram for the transition can be fully described based on measured macroscopic quantities of  the global system, without the necessity of considering the corresponding local uniform system as done with LDA. 

An additional insight into the physics of $\Pi$ can be gained by noticing the additional identity\cite{VRRBJP},
\begin{equation}
\Pi {\cal V} = \int \>p(\vec r) \>d^3 r ,\label{pploc}
\end{equation}
showing that, actually, the pressure parameter $\Pi$ is proportional to the integral of the local hydrostatic pressure $p(\vec r)$ over all space. To obtain this result we just need to remember that  mechanical equilibrium in the fluid is determined by the balance of local forces, namely,  $\nabla p(\vec r) + n(\vec r) \nabla V_{ext}(\vec r) = 0$. By using equation (\ref{piform}) the integration of the virial of these forces, $\vec r \cdot \vec f$, leads to (\ref{pploc}). A similar mechanical equilibrium argument has also been used by Horikoshi et al.\cite{Horikoshi}.

To measure the quantities involved in Eq.(\ref{piform}) across BEC transition we have used an experimental setup composed of a double MOT and a QUIC-type trap. The system has been described in detail elsewhere\cite{details-Brasil}. In brief, we collect $10^9$ $^{87}$Rb atoms in a MOT, producing a sample around $100 \mu$K. After a population transfer to the $|2,2\rangle$ hyperfine state, the sample is used to load a QUIC magnetic trap. While trapped, a RF-induced evaporation is applied to obtain quantum degeneracy. The trapping potential is harmonic with frequencies $ \omega_x = \omega_0$ and $\omega_y = \omega_z = 9 \omega_0$, where $\omega_0 = 2 \pi \times 23$ Hz. The characterization of the sample is done by absorption imaging it on a CCD camera after 15 ms time of flight. We typically produce BEC samples containing 1 - 7 $\times 10^5$ atoms.

Each acquired image is fitted with a bimodal distribution composed of a gaussian and an inverted parabola (Thomas-Fermi distribution) which represent, respectively, the thermal cloud and the condensate part.  By using the procedure described by Castin and Dum\cite{Castin}, one can obtain the ``in situ" density distribution of the cloud, $n(\vec r)$, which is cylindrically symmetric.
From those fittings, the temperature $T$ is obtained (gaussian tail), as well as the condensate fraction $N_0$ and the total number of atoms $N$ through a spatial integration. Finally, by using equation (\ref{piform}) we obtain the pressure parameter $\Pi$.

For the experiment here reported we have kept the trap frequencies fixed, that is ${\cal V} =$ constant. Temperature and number of atoms were varied by controlling the evaporative cooling final frequency and the time of evaporation\cite{details-Brasil} . We have performed about 300 runs of the experiments and arranged the results with similar number of particles for different temperatures.

\begin{figure}
\begin{center}
 \includegraphics[width=1.0\textwidth]{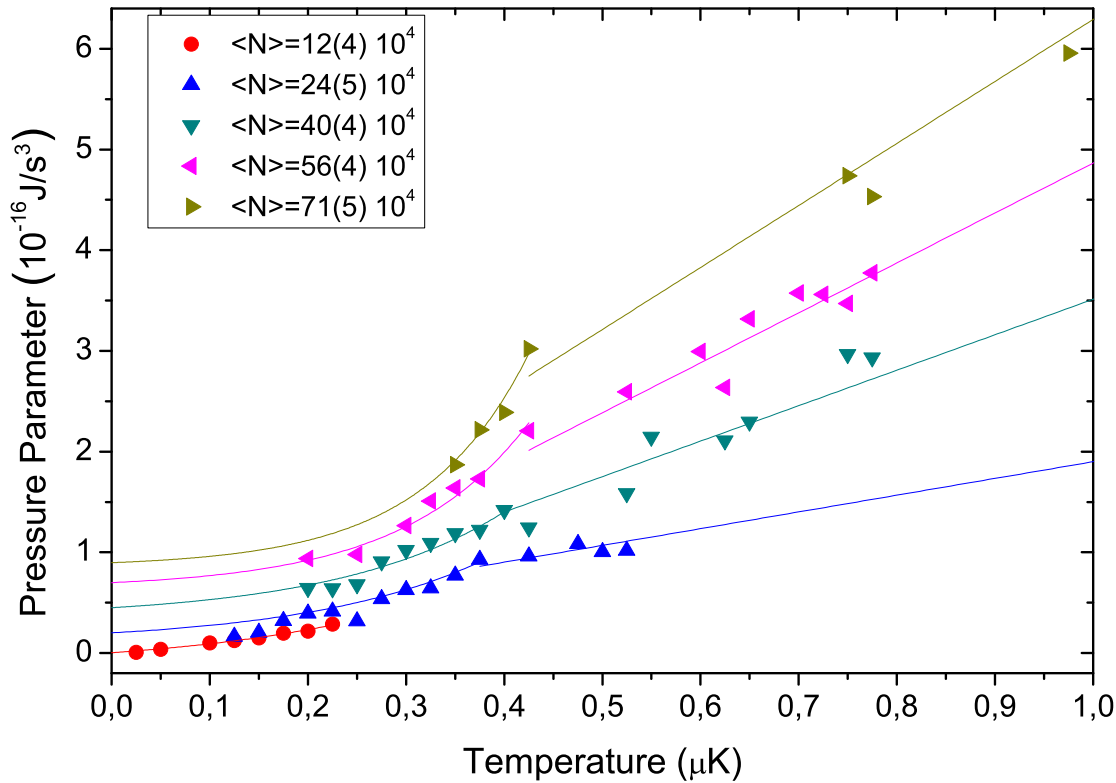}
\end{center}
\caption{
Pressure parameter $\Pi$ as a function of temperature $T$ for different constant number of particles $N$, crossing their corresponding critical temperature $T_c(N/{\cal V})$. Solid lines are best fittings through experimental points. Above the critical temperature the dependence is linear.}
\label{fig:pvst}
\end{figure}

Since we are expressing  $\Pi = \Pi(N/{\cal V},T)$, the experiment corresponds to the investigation of $\Pi$ vs $T$, while the density parameter $N/{\cal V}$ remains constant. The results of this experiment is shown in Fig.~\ref{fig:pvst}. For a fixed density parameter we observe two different regions of behavior as $T$ decreases. First, for high temperatures, $\Pi$ has an almost linear dependence both on $T$ and $N/{\cal V}$ (which is the prediction of an ideal Bose gas above BEC), until an abrupt change takes place, and the decrease of $\Pi$ with $T$ becomes faster than linear. This change of behavior characterizes the critical temperture $T_c = T_c(N/{\cal V})$ where BEC takes place. This change was also detected in the corresponding density profiles, that is, those below $T_c$ showed the characteristic peak of the condensate. The critical point $(\Pi_c,T_c)$ for each density parameter was determined by the crossing point of the extrapolations of both sides of the transition. Following the definition of $\Pi$ from equation (\ref{pi}), and the fact that for $T  > T_c$ the density clouds are very well fitted by a gaussian profile, the mentioned linear dependence is expected. That is, if one considers  the Bose distribution of particles in the non-interacting approximation,
\begin{equation}
\Pi \approx  \frac{Nk_B T}{\cal V} \frac{g_4(z)}{g_3(z)} \>\>\>\>{\rm(ideal \>\>gas}) ,
\end{equation}
with $g_n(z)$ the Bose function and $z = \exp(\mu/k_B T)$ the fugacity, the deviation from linearity in $T$ and $N/{\cal V}$ is rather small since $1 \ge g_4(z)/g_3(z) \ge 0.9$. Even if Hartree-Fock corrections are included\cite{VRRPRL2005}, the dependence of $\Pi$ on $T$ remains very close to linear for $T > T_c$. Below $T_c$, the behavior of $\Pi$ is intrinsically related to the superfluid nature of the gas, depending very strongly now on the magnitude of the interparticle interaction. That is,  if the interparticle interactions played no role, all the curves in Fig.~\ref{fig:pvst} would follow the same {\it ideal} transition line (similar to the dash line in Fig.~\ref{fig:transition}, see below).

From the critical points determined in Fig.~\ref{fig:pvst} one can extract the phase diagram $\Pi$ vs $T$ shown in Fig.~\ref{fig:transition}, where the critical line $\Pi_c$ vs $T_c$ separates the thermodynamic states into two domains, one where the fluid is thermal overall, and the other where there is a superposition of a Bose condensate fluid with a thermal component. The line $\Pi_c$ vs $T_c$ represents what is expected to be a continuous phase transition\cite{Yukalov} between a normal gas and a superfluid in an interacting Bose fluid confined by a harmonic trap. Below we address the issue of the order of the transition as predicted by Fig.~\ref{fig:pvst}. If the gas were ideal one should obtain $\Pi_c \sim T_c^4$ at the transition line, however, a logarithm plot reveals that a simple law $\Pi T^\gamma =$ constant is not obeyed for $\gamma$ constant. The exponent becomes larger as the temperature decreases. This feature, and those discussed below, are certainly due to the influence of interactions below $T_c$ and to the presence of the confining potential.

\begin{figure}
\begin{center}
 \includegraphics[width=1.0\textwidth]{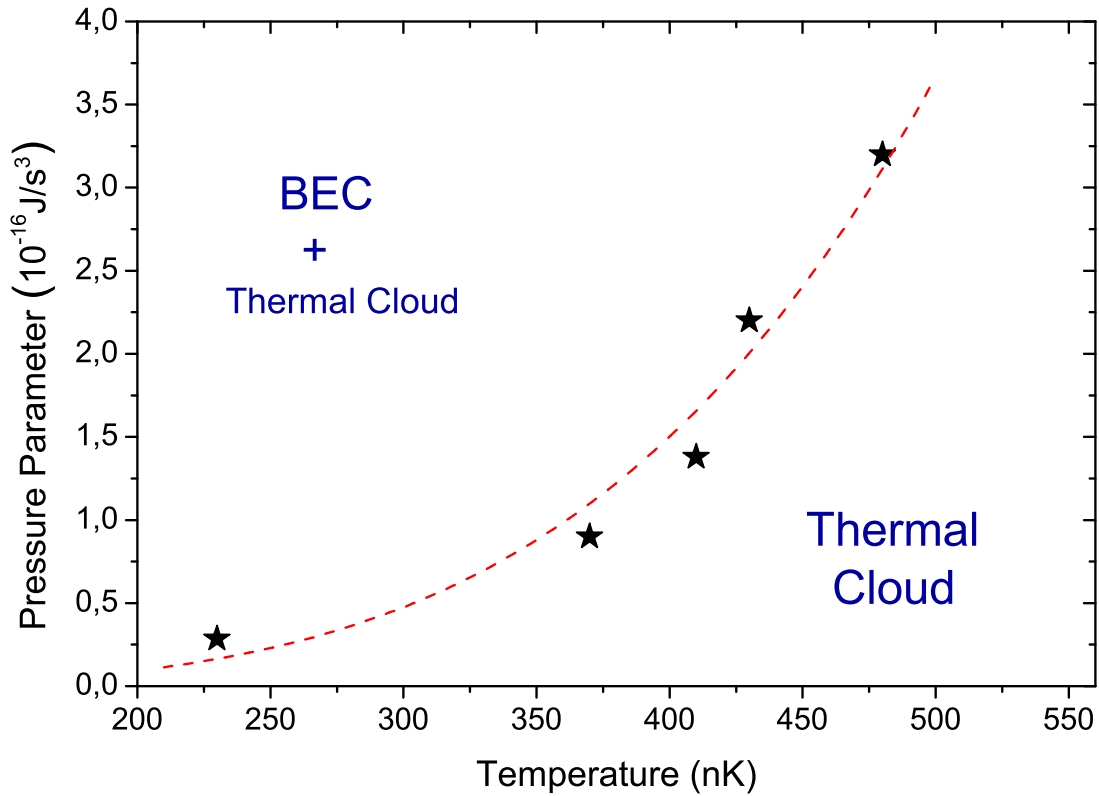}
\end{center}
\caption{
Phase diagram in terms of pressure parameter $\Pi$ and temperature $T$. The BEC transition line separates the two phases: purely thermal cloud and thermal cloud plus condensate.}
\label{fig:transition}
\end{figure}

\begin{figure}
\begin{center}
 \includegraphics[width=1.0\textwidth]{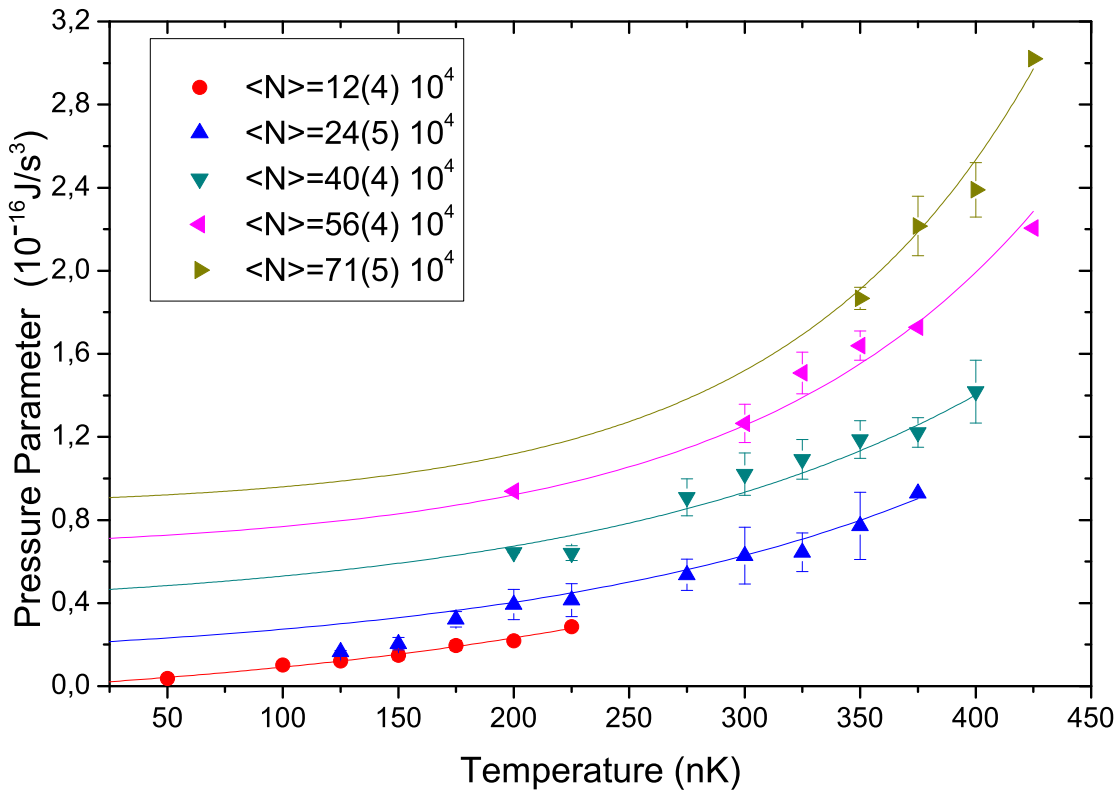}
\end{center}
\caption{Detail of pressure parameter dependence on temperature, for different fixed number of particles, $\Pi = \Pi(N/{\cal V},T)$, in the BEC region. The extrapolation of the experimental points allows for the determination of the pressure parameter at $T = 0$.}
\label{fig:ptzero}
\end{figure}

We now analyze data for  each isodensity curve (density parameter constant) shown in Fig.~\ref{fig:pvst}, for temperatures below $T_c$. These points are presented in Fig.~\ref{fig:ptzero}. Note that higher isodensity curves produce higher pressure parameter values at a given temperature. This appears as expected from an overall analysis of these variables in harmonic\cite{VRRPRL2005} and linear quadrupolar\cite{VRRPRE2008} traps using the Hartree-Fock approximation. The extrapolation of the experimental points in Fig.~\ref{fig:ptzero}, represented by the solid lines, allows the determination of $\Pi_0 = \Pi(N/{\cal V},T=0)$, the zero-temperature pressure parameter, which depends on the pure condensate density without the presence of the thermal component. At $T = 0$, the Thomas-Fermi (TF) approximation works on its best conditions.
Therefore, by using TF and LDA\cite{Dalfovo,Pethick} and assuming a cloud of $N$ interacting atoms with scattering length $a$, equation (\ref{piform}) at $T = 0$ yields,
\begin{equation}
\Pi_0 = \frac{1}{7} \left(15 \hbar^2 a \sqrt{m} \right)^{2/5} \left(N \omega_r^2 \omega_z\right)^{7/5} .\label{p0}
\end{equation}
It is found then, that at $T = 0$ the pressure parameter scales with the number of particles (keeping ${\cal V}$ fixed) as $\Pi_0 \sim N^\delta$ . A log-log plot of the values found from Fig.~\ref{fig:ptzero} yields the value $\delta = 1.5_{-0.2}^{+0.3}$, well within the expected value $1.4$ predicted by TF. The use of expression (\ref{p0}) together with the data presented in Fig.~\ref{fig:ptzero}, allow the determination of the scattering length, fully based on global thermodynamic quantities. However, this is not the scope of this paper.

As with standard pressure and volume in uniform fluid systems, the variables $\Pi$ and ${\cal V}$ provide us with a macroscopic thermodynamic description of the fluid without the necessity of relying on LDA. Because of the physical connection between the parameters $(\Pi,{\cal V})$ with $(p,V)$, the phase diagram reveals interesting macroscopic features not yet considered along the phase transition for the formation of a superfluid atomic phase in a trap.

As a first illustration, we note the striking similarity of the phase diagram of Fig.~\ref{fig:transition} with the normal gas to superfluid phase transition in the bottom part of the phase diagram $p-T$ of $^4$He, see Refs. \cite{He4a,He4b} for instance. We point out that this transition is different from the $\lambda$-transition which occurs between a normal {\it liquid} and a superfluid. In the gas-superfluid transition here seen the coexistence pressure parameter vanishes as the temperature goes to zero, just as in the analogous gas-superfluid transition diagram $p-T$ of $^4$He.  On the other hand, we also insist in the profound differences between considering a local picture and a global or thermodynamic one of a confined system. If studies were made of local variables only, the phase diagram obtained would refer to the bulk system and the transition line would not yield the information of the occurrence of BEC in the trapped gas. That is, knowledge of the bulk phase diagram solely does not allow a direct  calculation of the phase diagram of the inhomogeneous trapped gas. One needs first full knowledge of the number particle density $n = N/V$ and hydrostatic pressure $p$ of the homogeneous gas as a function of both chemical potential $\mu$ and temperature $T$, i.e. $n(\mu,T)$ and $p(\mu,T)$. Then, using LDA with the harmonic potential, a reconstruction of the phase diagram $\Pi$ vs $T$ such as in Fig.~\ref{fig:pvst} could be made. However, BEC occurs in the trapped inhomogeneous gas and this information is succinctly contained in the phase diagram of Fig.~\ref{fig:pvst} and Fig.~\ref{fig:transition}.

\begin{figure}
\begin{center}
 \includegraphics[width=1.0\textwidth]{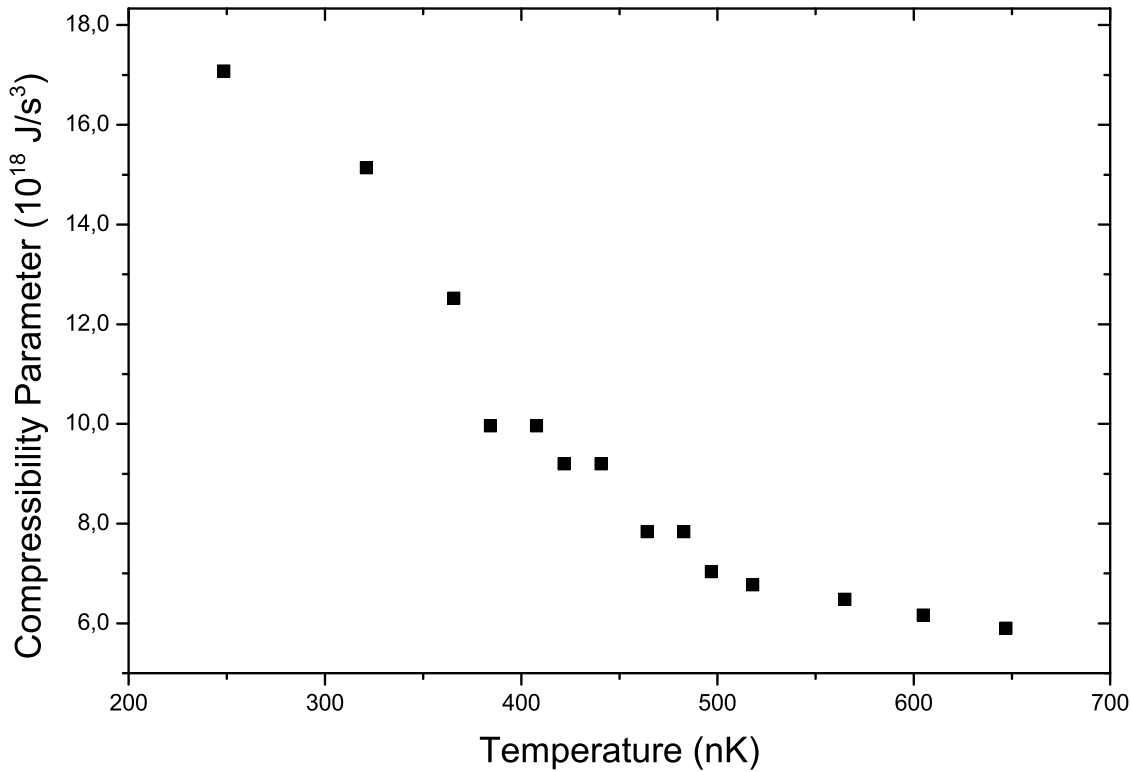}
\end{center}
\caption{
Isothermal compressibility parameter as a function of temperature for a fixed density parameter, $N = 2.8 \times 10^5 {\cal V}$. The location of the critical temperature corresponds to the change of the slope. This feature may signal a third-order phase transition.}
\label{fig:kt}
\end{figure}

The elucidation of the order of a phase transition is a very delicate issue that can only be resolved by either solving the problem exactly or by resorting to experimental evidence. The liquid to superfluid transition in $^4$He is a continuous second order one\cite{Landau} and the BEC transition in {\it bulk} systems is widely believed to be also of second order\cite{Lieb,Yukalov}. The features that determine this behavior is the divergence of the heat capacity and of the isothermal compressibility\cite{Ma}. The former is well exemplified by the $\lambda$-transition in $^4$He. Since the isothermal compressibility is related to the density or particle fluctuations, its divergence signals the appearance of critical behavior in the system\cite{Ma}. Here, by having knowledge of the equation of state $\Pi = \Pi(N/{\cal V},T)$ through the plots of Fig.~\ref{fig:pvst}, we can attempt to calculate the corresponding isothermal compressibility parameter $K_T$, which turns out to be related to the density fluctuations as well, that is,
\begin{eqnarray}
K_T &=& \frac{\cal V}{N} \left(\frac{\partial \Pi}{\partial N/{\cal V}}\right)_{T}^{-1} \nonumber \\
& = & \frac{\cal V}{N} \frac{1}{k_B T} \frac{\overline{\Delta N^2}}{N}
\end{eqnarray}
where the particle fluctuations are $\overline{\Delta N^2} = \overline{N^2} - {\overline N}^2$. The above expressions follow from the definition of $\Pi$, equation (\ref{pi}). From the data of Fig.~\ref{fig:pvst} we were able to obtain the isothermal compressibility for one isodensity curve ($N \approx 2.8 \times 10^5$) as a function of temperature. The result is shown in Fig.~\ref{fig:kt}, where we also show the approximated location of the corresponding BEC temperature $T_c$. Two features must be highlighted. First, we see no evidence that the isothermal compressibility parameter becomes very large at criticality (diverging in the thermodynamic limit) neither from above nor from below $T_c$, as expected in a second-order phase transition\cite{Ma}. But secondly and very interestingly, one sees a change of sign in the {\it curvature} of $K_T$ as a function of $T$, the zero curvature point nearly coinciding with the location of $T_c$. A plot of the {\it derivative} $(\partial K_T / \partial T)_{N/{\cal V}}$ will then show a pronounced (negative) peak at $T_c$. Although it is widely believed that phase transitions can only be of first or second order, this knowledge is solely based on uniform systems. Here, we advance the suggestion that BEC transition in a harmonically trapped gas may be of {\it third} order. We make this possible identification following Ehrenfest classification. That is, we identify that derivatives of the compressibility can be expressed as third-order derivatives of a free energy,
\begin{equation}
\left(\frac{\partial K_T}{\partial T}\right)_{N/{\cal V}} = - \frac{1}{{\cal V}} 
\left( \frac{\partial^3 G}{\partial T \partial \Pi^2} + \left(\frac{\partial \Pi}{\partial T} \right)_{N/{\cal V}}  \frac{\partial^3 G}{\partial \Pi^3} \right)
\end{equation}
where $G=G(N,\Pi,T)$ is Gibbs free energy. If the peak of $(\partial K_T / \partial T)_{N/{\cal V}}$ at $T_c$ becomes more and more pronounced, as
the system becomes larger (i.e. approaching the thermodynamic limit), it would indicate the divergence of a third derivative of the free energy. This would also indicate that while the density fluctuations do not diverge, their derivatives do so. This, to the best of our knowledge, is a property never reported before. We believe that this behavior is due to the fact that the system is confined by a harmonic potential and that it is not uniform. 
This behavior can be seen to be already present as a precursor in the ideal  Bose gas in a harmonic trap. That is, one can show that for  $T \ge T_c$, it  is obeyed,
\begin{equation}
\left(\frac{\partial K_T^{ideal}}{\partial T}\right)_{N/{\cal V}} = k_B \left(\frac{\cal V}{N}\right)^2 \frac{k_B T_c}{\hbar^3 g_2(z)} 
\left[2 g_2(z)g_2(z) - 3 g_3(z)g_1(z)\right] .
\end{equation}
Approaching BEC from above for $N/{\cal V}$ fixed, i.e. $z \to 1^+$, one finds that both, $g_3(1) = \zeta(3)$ and $g_2(1) = \pi^2/6$, are finite, while $g_1(1) \to \infty$. Thus $(\partial K_T^{ideal}/\partial T)_{N/{\cal V}} \to - \infty$ as the experiments suggest. 
The full elucidation of this aspect requires a much thorough analysis of the equation of state and will be reported elsewhere.

The capability to describe an inhomogeneous quantum gas via macroscopic parameters creates means to access other thermodynamic quantities proper of the trapped gas that cannot be accessed using local variables within LDA. For instance, the heat capacity at constant ${\cal V}$, $C_{\cal V}$, is a quantity that pertains to the full macroscopic system and cannot be directly obtained from direct integration of local quantities\cite{VRRPRE2008}. It can be shown that the heat capacity may be obtained by full knowledge of the equation of state $\Pi = \Pi(N/{\cal V},T)$ and by independent measurements of the adiabatic change of the temperature as the volume parameter is changed\cite{VRRPRE2008}. Further, as done in this paper, one can also calculate the corresponding isothermal compressibility parameter $K_T \equiv (N/{\cal V}) (\partial \Pi/\partial (N/{\cal V}))_T$. The elucidation of these quantities,  $K_T$ and $C_{\cal V}$, through BEC, together with the  phase diagram here shown, would reveal the critical details of interacting BEC in a trap, a question still unanswered to this day.

\ack{We appreciate collaboration and helpfull discussions with J. Seman, P. Castilho, P.Tavarez, S.R. Muniz and G. Telles. The experimental work has received financial support from FAPESP, CNPq and CAPES, Brazil. VRR acknowledges support from DGAPA UNAM IN-116110.}

\end{document}